\documentclass{article}
\usepackage{graphicx}
\usepackage{amssymb}
\usepackage{url}
\usepackage{epsfig}
\usepackage{amsfonts,amssymb,amsbsy}
\usepackage{psfrag}
\usepackage{amsmath}
\usepackage{comment}

\def\e{\begin{equation}} 
\def\f{\end{equation}} 

\def\##1{{\mbox{\textbf{#1}}}}
\def\%#1{{\mbox{\boldmath $#1$}}}

\def\=#1{{\overline{\overline{\mathsf #1}}}}

\def\/{\over}
\def\*{^{\displaystyle*}}

\def\.{\cdot}
\def\x{\times}
\def\:{\over}
\def\oo{\infty}

\def\D{\nabla}
\def\d{\partial}

\def\ra{\rightarrow}

\def\Ra{\Rightarrow}

\def\l#1{\label{eq:#1}}
\def\r#1{(\ref{eq:#1})}
\def\am{\left(\begin{array}{c}}
\def\amm{\left(\begin{array}{cc}}
\def\ammm{\left(\begin{array}{ccc}}
\def\ammmm{\left(\begin{array}{cccc}}
\def\a{\end{array}\right)}

\def\add{\left|\begin{array}{cc}}
\def\addd{\left|\begin{array}{ccc}}
\def\adddd{\left|\begin{array}{cccc}}
\def\ad{\end{array}\right|}

\def\E{\epsilon}

\def\h{\eta}

\def\la{\lambda}

\def\M{\mu}
\def\o{\omega}

\def\t{\tau}

\def\TH{\theta}

\def\VF{\varphi}

\def\bcal#1{{\mbox{\boldmath $\cal#1$}}}

\def\det{{\rm det}}

\def\�#1{\underline{\bf #1\mit}}

\begin{document}

\title{Realization of a Spherical Boundary by a Layer of Wave-Guiding Medium}
\author{Ismo V. Lindell, Johannes Markkanen,\\ Ari~Sihvola, and Pasi Yl\"a-Oijala}

\date{Aalto University School of Electrical Engineering\\ Department
  of Radio Science and Engineering\\ Box 13000, FI--00076 AALTO, Finland\\ \today}

\maketitle

\begin{abstract}
In this paper the concept of wave-guiding medium, previously introduced for planar structures, is defined for the spherically symmetric case. It is shown that a quarter-wavelength layer of such a medium serves as a transformer of boundary conditions between two spherical interfaces. As an application, the D'B'-boundary condition, requiring vanishing of normal derivatives of the normal components of D and B field vectors, is realized by transforming the DB-boundary conditions. To test the theory, scattering from a spherical DB object covered by a layer of wave-guiding material is compared to the corresponding scattering from an ideal D'B' sphere, for varying medium parameters of the layer. \end{abstract}

\section{Introduction}

Electromagnetic problems are often mathematically defined in terms of differential equations and boundary conditions which appear on surfaces limiting the region of interest. Here we may separate two different ways to connect mathematical boundary conditions and a physical structure. In the analytical way one starts from a given physical problem involving a structure of electromagnetic media and tries to replace existing medium interfaces by boundary conditions for the purpose of simplifying the mathematical problem. In such cases the boundary conditions are approximative. In the synthetic way a desired structure is mathematically designed in terms of boundary conditions, which raises the question how to realize them in terms of physical media. This latter way is of concern here. 

In \cite{AP07} it was shown that any given impedance boundary, defined by boundary conditions of the type
\e \#u_z\x\#E = \=Z_s\.\#H \f
on a planar surface $z=0$ with unit normal $\#u_z$, can be realized by an interface of a wave-guiding medium defined by medium dyadics of the form
\e \=\E = \E_z\#u_z\#u_z + \=\E_t, \l{E}\f
\e \=\M = \M_z\#u_z\#u_z + \=\M_t, \l{M}\f
when the axial parameters grow without limit as $\E_z\ra\oo$, $\M_z\ra\oo$. It was shown that the dyadics $\=\E_t$ and $\=\M_t$ transverse to the medium axis can be designed to realize the surface impedance dyadic $\=Z_s$. 

As another example, the DB and D'B' boundary conditions on a plane \cite{DB,AP10}, respectively defined in terms of normal field components as
\e \#u_z\.\#D=0,\ \ \ \ \ \#u_z\.\#B=0, \f
and 
\e \d_z(\#u_z\.\#D)=0,\ \ \ \ \ \d_z(\#u_z\.\#B)=0, \f
have been shown to be realizable in terms of physical structures. In fact, already in 1959, it was found that the DB boundary can be realized by the interface of a uniaxially anisotropic medium \r{E} and \r{M} satisfying $\E_z\ra0$ and $\M_z\ra0$ \cite{Rumsey}. The DB medium has recently found application in the design of cloaking structures \cite{Kong,Yaghjian,Weder,Arthur}.

For the planar D'B' boundary, originally introduced in \cite{AP10}, a realization was found only recently. In \cite{EM16} it was shown that a layer of wave-guiding medium, called the quarter-wave transformer, upon a planar DB boundary produces D'B' boundary conditions at its interface.

The purpose of the present paper is to study the properties of the wave-guiding medium of spherical symmetry in order to extend the previous results from planar boundaries to spherical boundaries. The DB and D'B' conditions for the spherical boundary take the respective form  \cite{AP10}
\e \#u_r\.\#D=D_r=0,\f
\e \#u_r\.\#B=B_r=0, \f
and 
\e \D\.(\#u_r\#u_r\.\#D)= \d_r(r^2D_r)=0, \l{D'}\f
\e \D\.(\#u_r\#u_r\.\#B)= \d_r(r^2B_r)=0. \l{B'}\f
It was shown in \cite{AP09} that objects with either DB or D'B' boundary conditions with a certain symmetry in their geometry have no backscattering, i.e., they are invisible for the monostatic radar.

\section{Fields in spherical wave-guiding medium}

Let us consider the spherically symmetric anisotropic medium defined by
\e \=\E = \E_r\#u_r\#u_r + \E_t(\=I-\#u_r\#u_r), \f
\e \=\M = \M_r\#u_r\#u_r + \M_t(\=I-\#u_r\#u_r), \f
where $\E_r,\E_t,\M_r$ and $\M_t$ are constant parameters. The Maxwell equations can be represented in spherical coordinates as \cite{Bladel}
\e \frac{1}{r^2\sin\TH}\det\ammm \#u_r & r\#u_\TH & r\sin\TH\#u_\VF\\ \d_r & \d_\TH & \d_\VF\\ E_r & rE_\TH & r\sin\TH E_\VF\a = -j\o\=\M\.\#H, \f
\e \frac{1}{r^2\sin\TH}\det\ammm \#u_r & r\#u_\TH & r\sin\TH\#u_\VF\\ \d_r & \d_\TH & \d_\VF\\ H_r & rH_\TH & r\sin\TH H_\VF\a = j\o\=\E\.\#E. \f
Expressing the fields in their radial and transverse components as
\e \#E = \frac{\#u_r}{\E_r}D_r + \#u_\TH E_\TH+ \#u_\VF E_\VF,\f
\e \#H = \frac{\#u_r}{\M_r}B_r + \#u_\TH H_\TH+ \#u_\VF H_\VF,\f
and assuming that the radial components of the medium parameters satisfy the wave-guiding medium condition
\e \E_r\ra\oo,\ \ \ \ \M_r\ra\oo, \f
we have $E_r\ra0$ and $H_r\ra0$, whence the Maxwell equations for the transverse field components are simplified to
\e \d_r(r E_\VF) = jk_t\h_t rH_\TH, \f
\e \d_r(r E_\TH) = -jk_t\h_t rH_\VF, \f
\e \h_t\d_r(rH_\VF) = -jk_t rE_\TH, \f
\e \h_t\d_r(rH_\TH) = jk_t rE_\VF, \f
where we denote
\e k_t = \o\sqrt{\M_t\E_t},\ \ \  \ \h_t = \sqrt{\M_t/\E_t}. \f
After elimination of fields, the equations become
\e (\d_r^2 + k_t^2) rE_\VF=0, \l{1}\f
\e (\d_r^2 + k_t^2) rE_\TH=0, \l{2}\f
\e (\d_r^2 + k_t^2) rH_\VF=0, \l{3}\f
\e (\d_r^2 + k_t^2) rH_\VF=0, \l{4}\f
It is noteworthy that there is no restriction to the $\TH$ and $\VF$ dependence for the fields. Thus, the four field components satisfying \r{1} -- \r{4} can be expressed in the form 
\e F(\#r) = F_+(\TH,\VF)e^{-jk_tr} + F_-(\TH,\VF)e^{jk_tr}. \f
Let us omit the dependence on $\TH$ and $\VF$ in the notation. 

After solving the transverse field components, the radial fields $B_r$ and $D_r$ can be obtained from the Maxwell equations as
\e B_r= \frac{j}{\o r\sin\TH}(\d_\TH(\sin\TH E_\VF) - \d_\VF E_\TH), \l{Br}\f
\e D_r=-\frac{j}{\o r\sin\TH}(\d_\TH(\sin\TH H_\VF) - \d_\VF H_\TH). \l{Dr}\f

\section{Field solutions}

The solutions for the transverse field components can be expressed as
\e E_\VF(r) = \frac{1}{r}(A_+e^{-jk_tr}+ A_-e^{jk_tr}), \l{EVF}\f 
\e \h_tH_\TH(r) = -\frac{1}{r}(A_+e^{-jk_tr}- A_-e^{jk_tr}), \f 
\e E_\TH(r) = \frac{1}{r}(C_+e^{-jk_tr}+ C_-e^{jk_tr}), \f 
\e \h_tH_\VF(r) = \frac{1}{r}(C_+e^{-jk_tr}- C_-e^{jk_tr}), \l{HVF}\f
where the coefficients $A_\pm, C_\pm$ are functions of $\TH$ and $\VF$. The radial fields can be obtained by substituting these in \r{Br} and \r{Dr}. In view of the D'B'-boundary conditions \r{D'} and \r{B'}, let us define modified radial field quantities as
$$ B_m(r) = r^2B_r(r)$$
\e = \frac{j}{\o \sin\TH}(\d_\TH(\sin\TH (rE_\VF)) - \d_\VF (rE_\TH)),\f
$$ D_m(r)=r^2D_r(r)$$
\e =-\frac{j}{\o \sin\TH}(\d_\TH(\sin\TH (rH_\VF)) - \d_\VF (rH_\TH)). \f
The modified fields have the general form
\e B_m(r) = B_+ e^{-jk_tr} + B_-e^{jk_tr}, \l{Bm}\f
\e D_m(r) = D_+e^{-jk_tr} + D_-e^{jk_tr}, \l{Dm}\f
with coefficients depending on those of the transverse fields as
\e B_\pm = \frac{j}{\o \sin\TH }(\d_\TH(\sin\TH A_\pm) - \d_\VF C_\pm), \f
\e D_\pm = \mp\frac{j}{\o \h_t\sin\TH }(\d_\TH(\sin\TH C_\pm) + \d_\VF A_\pm). \f

The quantities $B_m,D_m$ propagate radially for any transverse coordinate $\TH,\VF$ just like waves in a linear transmission line. Forming two new functions involving radial derivatives as
\e B_m'(r) = \frac{j}{k_t}\d_rB_m = B_+ e^{-jk_tr} - B_-e^{jk_tr}, \l{Bm'}\f
\e D_m'(r) = \frac{j}{k_t}\d_rD_m = D_+ e^{-jk_tr} - D_-e^{jk_tr}, \l{Dm'}\f
the coefficents $B_\pm,D_\pm$ can be solved from \r{Bm}, \r{Dm}, \r{Bm'} and \r{Dm'} as
\e B_\pm = \frac{1}{2}(B_m(r)\pm B_m'(r))e^{\pm jk_tr}, \f
\e D_\pm = \frac{1}{2}(D_m(r)\pm D_m'(r))e^{\pm jk_tr}. \f

Let us now consider the region between two spherical surfaces, $a\leq r\leq b$. Assuming values of $B_m,D_m,B_m'$ and $D_m'$ known on $r=a$ we can find their values on $r=b$ as
\e \am B_m(b)\\ B_m'(b)\a =  \cal{K} \am B_m(a)\\ B_m'(a)\a, \f
\e \am D_m(b)\\ D_m'(b)\a =  \cal{K} \am D_m(a)\\ D_m'(a)\a, \f
through the matrix
\e \cal{K} = \amm \cos k_t(b-a)  & -j\sin k_t(b-a) \\ -j\sin k_t(b-a) & \cos k_t(b-a)\a. \f

Similarly, we can find the following relations between the transverse field components,
\e \am bE_\VF(b)\\ b\h_tH_\TH(b)\a = {\cal L}\am aE_\VF(a)\\ a\h_t H_\TH(a)\a, \f
\e \am bE_\TH(b)\\ b\h_tH_\VF(b)\a = {\cal K}\am aE_\TH(a)\\ a\h_t H_\VF(a) \a, \f
with
\e \cal{L} = \amm \cos k_t(b-a)  & j\sin k_t(b-a) \\ j\sin k_t(b-a) & \cos k_t(b-a)\a. \f

\section{The quarter-wave transformer}

Let us now assume that the distance between the two spherical surfaces $b-a$ satisfies
\e k_t(b-a)=\pi/2, \f
whence $\cos k_t(b-a)=0$ and $\sin k_t(b-a)=1$. Defining the wavelength $\la_t$ by $k_t\la_t=2\pi$, we have $b-a=\la_t/4$. In this case the layer of wave-guiding medium has quarter wavelength thickness and the fields at the two surfaces obey the relations
\e B_m(b) = -jB_m'(a),\ \ \ \ B_m'(b) = -jB_m(a), \f
\e D_m(b) = -jD_m'(a),\ \ \ \ D_m'(b) = -jD_m(a), \f
\e bE_\VF(b) = ja\h_t H_\TH(a),\ \ \ \ b\h_tH_\TH(b) = jaE_\VF(a), \f
\e bE_\TH(b) = -ja\h_t H_\VF(a),\ \ \ \ b\h_tH_\VF(b) = -jaE_\TH(a). \f
The quarter-wavelength layer serves as a transformer of boundary conditions on $r=a$ to other boundary conditions on $r=b$. Let us list a few examples.

\begin{itemize}
\item DB-boundary to D'B'-boundary  
\e B_m(a)=0\ \ \ \Ra\ \ \ B_m'(b)=0, \f
\e D_m(a)=0\ \ \ \Ra\ \ \ D_m'(b)=0. \f
\item D'B'-boundary to DB-boundary  
\e B_m'(a)=0\ \ \ \Ra\ \ \ B_m(b)=0, \f
\e D_m'(a)=0\ \ \ \Ra\ \ \ D_m(b)=0. \f
\item PEC boundary to PMC boundary
\e E_\VF(a)=0\ \ \ \Ra\ \ \ H_\TH(b)=0, \f
\e E_\TH(a)=0\ \ \ \Ra\ \ \ H_\VF(b)=0.\f
\item PMC boundary to PEC boundary
\e H_\VF(a)=0\ \ \ \Ra\ \ \ E_\TH(b)=0, \f
\e H_\TH(a)=0\ \ \ \Ra\ \ \ E_\VF(b)=0.\f
\item PEMC boundary to PEMC boundary \cite{PEMC}
\e H_\VF(a)+ME_\VF(a)=0\ \ \ \Ra\ \ \ H_\TH(b)- \frac{1}{M\h_t^2}E_\TH(b)=0, \f
\e H_\TH(a)+ME_\TH(a)=0\ \ \ \Ra\ \ \ H_\VF(b)- \frac{1}{M\h_t^2}E_\VF(b)=0. \f
\end{itemize} 

The spherical quarter-wave transformer is quite similar to the previously studied planar quarter-wave transformer \cite{EM16}. In particular, the D'B'-boundary conditions can be realized on the spherical surface by applying the transformer layer upon a DB boundary whose realization is previously known.

Because the parameters $\E_t$ and $\M_t$ may be freely chosen, the distance between the two spherical surfaces can be made as small as we wish by choosing large values for $\E_t,\M_t$. However, we must take care that the parameters $\E_r,\M_r$ must be larger by an order of magnitude, because $\E_t/\E_r$ and $\M_t/\M_r$ must be small. When these conditions are met, the transformer layer can be made a thin sheet on the sphere of radius $a$.

\section{Numerical examples}

As an example, let us consider plane-wave scattering from a layered
sphere simulating the sphere on which a D'B' boundary is forced. The
sphere is located in free space defined by parameters $\E_o,\M_o$. In
the following, we compare the scattering behavior of an ideal D'B'
sphere and its different material realizations.

\subsection{Volume integral equations}

The scattering fields are found by first solving the equivalent polarization currents in the whole sphere (region $V$) from integral equations. The total time-harmonic electric $\#E$ and magnetic $\#H$ fields can be expressed via the volume equivalence principle \cite{CJMS} as
\e \#E(\#r) = \#E^{inc}(\#r) + \frac{1}{j\o\E_o}(\D\D+k_o^2\=I)\.\bcal{S}(\#J) -\D\x\bcal{S}(\#M), \l{JvMvRep1}\f
\e \#H(\#r) = \#H^{inc}(\#r) +\D\x\bcal{S}(\#J) + \frac{1}{j\o\M_o}(\D\D+k_o^2\=I)\.\bcal{S}(\#M), \l{JvMvRep2}\f
where $\#E^{inc}$ and $\#H^{inc}$ are the incident electric and magnetic fields, and $k_o = \o\sqrt{\E_o \M_o}$. The volume integral operator in \r{JvMvRep1} and \r{JvMvRep2} is defined by 

\e {\bcal S}(\#F) = \int\limits_V G_o(\#r,\#r')\,\#F(\#r')\,dV',  \l{Sope} \f
where $G_o$ is the free-space Green function. The equivalent electric and magnetic polarization currents are defined as 
\begin{equation}
\begin{array}{ccc}
  \#J(\#r) &=& j\o(\=\E(\#r)-\E_o\=I)\.\#E(\#r)
   \\ [2mm]
  \#M(\#r) &=& j\o(\=\M(\#r)-\M_o \=I)\.\#H(\#r). \\
\end{array} \label{JVMV}
\end{equation}

By applying the volume equivalence principle, definitions of the volume equivalent currents $\#J$ and $\#M$, vector identities, and the fact that the operator \r{Sope} satisfies the Helmholtz equation, we obtain the following volume integral equation formulation,  
\e -j\o\E_o\=\t_E\.\#E^{inc}(\#r) = -\=\E\.\#J + \=\t_E\.(\D\x\D\x\bcal{S}(\#J)) - j\o\E_o\=\t_E\.\D\x{\bcal{S}}(\#M), \l{JM1}\f
\e -j\o\M_o\=\t_M\.\#H^{inc}(\#r) = -\=\M\.\#J + j\o\M_o\=\t_M\.\D\x\bcal{S}(\#J) + \=\t_M\.(\D\x\D\x\bcal{S}(\#M)), \l{JM2}\f
with the contrast dyadics defined by
\e \E_o\=\t_{E} = \=\E-\E_o\=I,\ \ \ \ \ \M_o\=\t_{M} = \=\M-\M_o\=I.\f

The formulation \r{JM1}, \r{JM2} can be discretized using Galerkin's method with piecewise constant vector basis and testing functions. The hypersingularity of the kernel is reduced by moving one derivative into the testing function through integrating by parts. The remaining derivatives are then moved to operate on the Green function.

\subsection{Scattering from a layered sphere}

\begin{figure}[htb]
\centering
\psfrag{db}[c][0][1][0] {$\E = \M = 0$}
\psfrag{er}[c][0][1][0] {$\E_r = \M_r = \infty$}
\psfrag{et}[c][0][1][0] {$\E_t/\E_o = \M_t/\M_o = 2$}
\psfrag{et2}[c][0][1][0] {$\E_t/\E_o=\M_t/\M_o= 1$}
\psfrag{a1}[c][0][1][0] {$0.2$\,m}
\psfrag{a2}[c][0][1][0] {$0.8$\,m}
\psfrag{a3}[c][0][1][0] {$0.6$\,m}
\psfrag{a4}[c][0][1][0] {$0.4$\,m}
\includegraphics[width=0.9\textwidth]{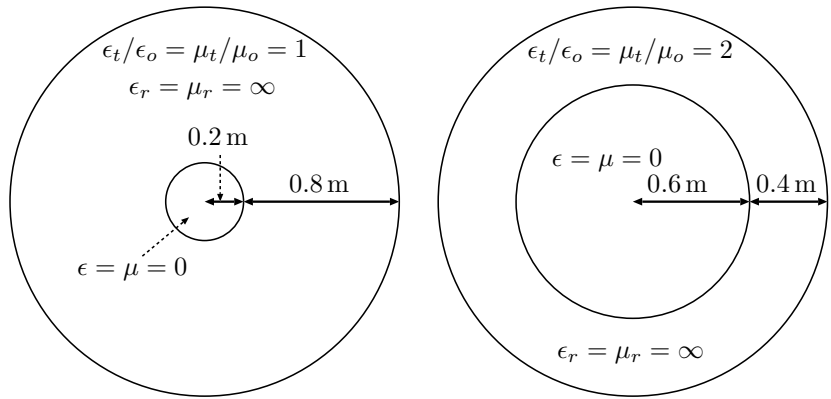}
\caption{\label{fig_geom} D'B'-boundary conditions on a spherical
  surface of radius $b=1$\, m can be realized by a quarter-wave
  transformer layer upon a sphere of radius $a$ with DB-boundary
  conditions. The thickness $b-a$ of the layer can be controlled by
  the components of permittivity and permeability transverse to the
  radial direction in the layer, $\E_t,\M_t$. The two cases are
  considered here as examples.}
\end{figure}

Let us now consider scattering from a layered sphere of radius
$b=1$\,m, for an incident plane wave with free-space wavelength $\la_o
= 3.2$\,m, propagating in the direction of positive $z$ axis
($\TH=0$). Fig.\ \ref{fig_geom} presents two possible realizations for
the D'B' sphere with transverse parameters either $\M_t=\M_o$,
$\E_t=\E_o$ or $\M_t=2\M_o$, $\E_t=2\E_o$.  In order to obtain DB
conditions at the surface $r=a$, the permittivity and permeability of
the inner sphere are set to zero \cite{Istanbul}. In Fig.\
\ref{fig_geom} (left) the thickness of the quarter-wave transformer
layer is $b-a=0.8$\,m since $\la_t=\la_o = 3.2$\,m. In Fig.\
\ref{fig_geom} (right), the reduced wavelength becomes $\la_t = \la_o
/2 = 1.6$\,m and therefore the layer thickness equals $b-a=0.4$\,m.

%\begin{figure}
%	\centering
%		\includegraphics[width=0.9\textwidth]{pallot.eps}
%\end{figure}

\begin{figure}[htb]
\centering
\includegraphics[width=0.9\textwidth]{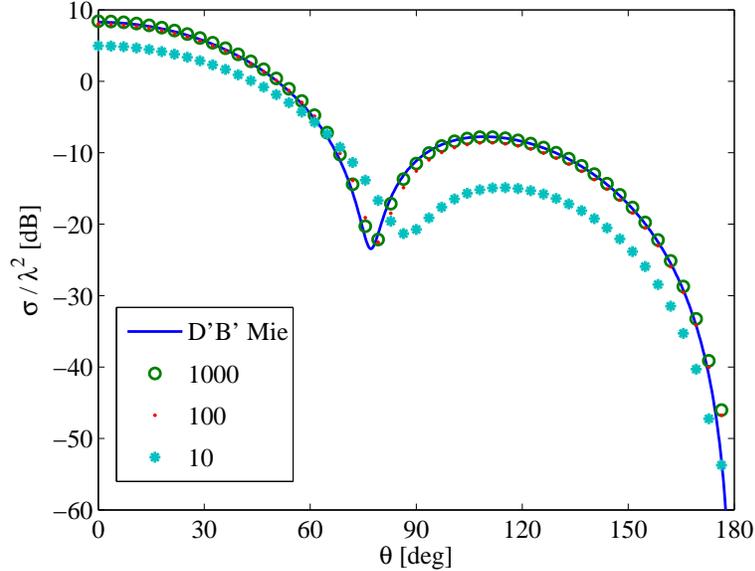}
\caption{Calculated bistatic scattering cross section corresponding to
  the realization of the D'B' sphere. Different values of radial
  permittivity and permeability $\E_r/\E_o=\M_r/\M_o$ are used to
  approximate the quarter-wave transformer while the relative
  tangential components are $\E_t=\E_o$, $\M_t=\M_o$. The ideal D'B'
  scattering is computed using Mie series.  The geometry of the
  realization can be seen in Fig.\ \ref{fig_geom} (left).}
\label{fig_rcs}
\end{figure}
\begin{figure}[htb]
\centering
\includegraphics[width=0.9\textwidth]{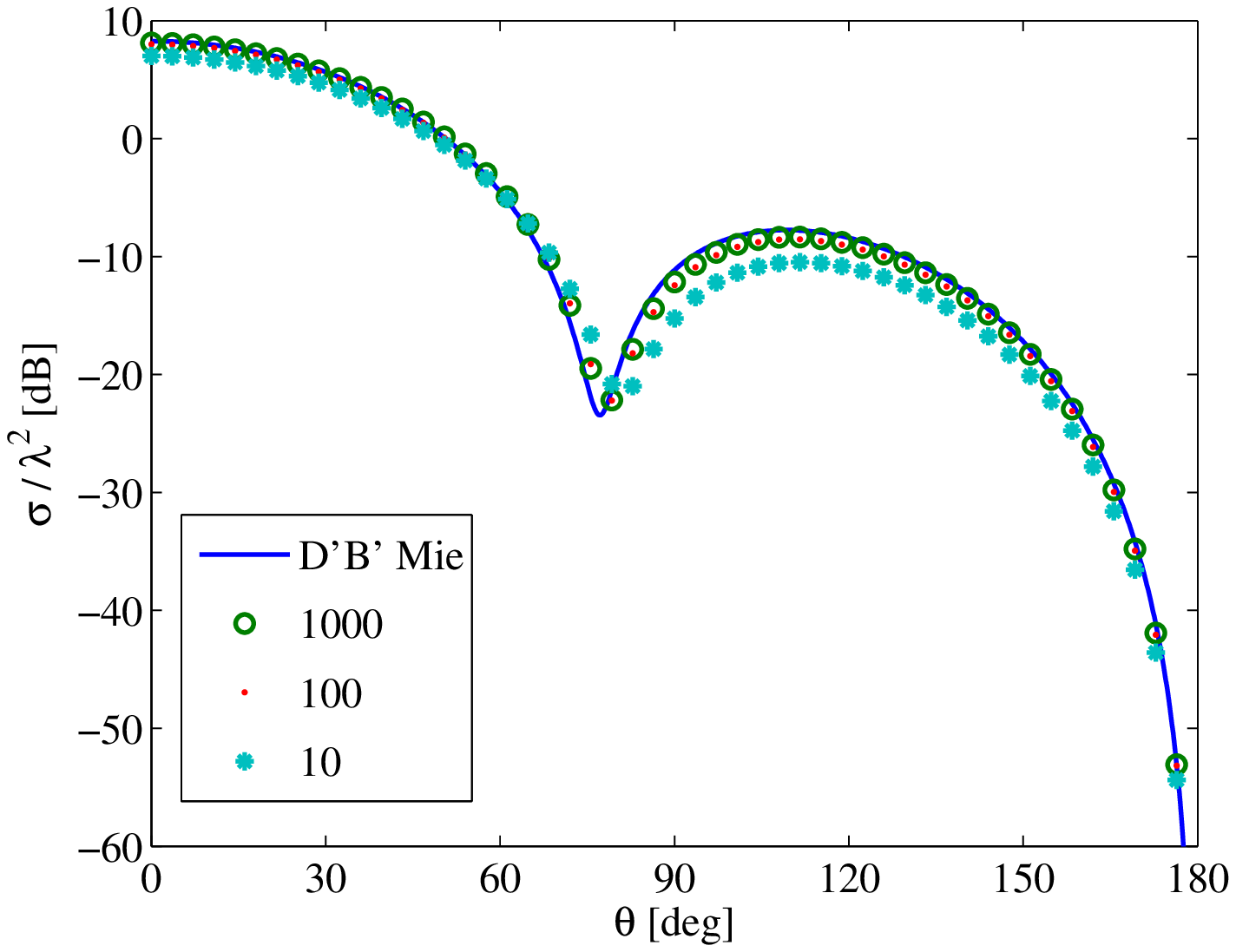}
\caption{Calculated scattering cross section of the realization of the
  D'B' sphere with $b-a=0.4$ m, corresponding to Fig.\ \ref{fig_geom}
  (right) for different values of radial permittivity and permeability
  $\E_r/\E_o=\M_r/\M_o$.}
\label{fig_rcs2}
\end{figure}

Fig.~\ref{fig_rcs} shows the scattering cross section (SCS) of the
coated sphere in Fig.~\ref{fig_geom} (left), as calculated by the
volume integral equation method. The calculated SCS of the sphere of
Fig.~\ref{fig_geom} (right) is presented in Fig.~\ref{fig_rcs2}. We
have used different values of radial permittivity and permeability in
order to get an idea on how large radial components of permittivity
and permeability are required in approximating the D'B' boundary. Note
that, due to the self-dual character of the structure, the scattering
pattern is rotationally symmetric, i.e.\ E and H planes have the same
patterns.

Clearly, radial parameter values $\E_r=10\E_o$, $\M_r=10\M_o$ yield a
poor approximation to the quarter-wave transformer, whence the sphere
markedly deviates from one with the D'B' boundary. However, for larger
values of radial permittivity and permeability, the scattering cross
sections of the two coated spheres approach to that of the ideal D'B'
sphere. The ideal D'B' scattering is computed from an exact Mie
scattering solution \cite{AP09}.  From the numerical results it
appears that, for larger values for the transverse parameters
$\E_t,\M_t$, lower values of the radial parameters $\E_r,\M_r$ can be
accepted for the realization. Furthermore, one may note that, for all
parameter values used in the computation, the layered sphere has very
small scattering in the backward direction ($\TH=180^\circ$). This is
in accord with the theory stating that both DB and D'B' spheres are
invisible for the monostatic radar \cite{AP09}.

\section{Conclusion}

An anisotropic medium with infinitely large axial permittivity and
permeability has been called waveguiding medium and its properties in
transforming impedance-boundary conditions between two planar
interfaces has been studied in \cite{AP07}. In \cite{EM16} it was
shown that a layer of waveguiding medium can be also used to transform
between planar DB and D'B' boundary condition which involve vanishing
of normal components of the $\#D$ and $\#B$ fields or their normal
derivatives. In the present paper the spherical waveguiding medium is
defined and its transforming properties are studied. In particular, it
is shown that a quarter-wavelength layer of waveguiding medium can be
used to transform a DB boudary to a D'B' boundary. Since it is known
that a spherical DB boundary can be realized by a medium with
vanishing permittivity and permeability \cite{Istanbul}, this gives a
means to realize a spherical D'B' boundary which until now has had no
realization whatever. The theory is verified numerically through
volume-integral equation approach by considering scattering from a
layered spherical object and comparing with that from an ideal D'B'
sphere. It is seen that the realization approaches the ideal case when
the radial parameters of the quarter-wavelength layer grow large. Such
an object is of interest because it has zero backscattering like the
corresponding DB sphere \cite{AP09}.

\end{document}